\documentclass[twocolumn,secnumarabic,amssymb, nofootinbib, aps, prd]{revtex4-1}
\usepackage{graphicx}
\usepackage{amsmath}
\renewcommand{\theequation}{\arabic{section}.\arabic{equation}}

\newcommand{\classoption}[1]{\texttt{#1}}
\newcommand{\macro}[1]{\texttt{\textbackslash#1}}
\newcommand{\m}[1]{\macro{#1}}
\newcommand{\env}[1]{\texttt{#1}}
\def\be{\begin{equation}}
\def\ee{\end{equation}}
\def\bq{\begin{equation}}
\def\eq{\end{equation}}
\def\bqa{\begin{eqnarray}}
\def\eqa{\end{eqnarray}}
\def\roughly#1{\mathrel{\raise.3ex
\hbox{$#1$\kern-.75em\lower1ex\hbox{$\sim$}}}}
\def\lsim{\roughly<}
\def\gsim{\roughly>}

\begin{document}



\title{Color Dipole Picture at Low-x DIS:\\
The Mass Range of Active Photon Fluctuations}%

\author{Masaaki Kuroda}
\email{kurodam@law.meijigakuin.ac.jp}
\affiliation{Center for Liberal Arts, Meijigakuin University,
Yokohama, Japan}
\author{Dieter Schildknecht}%
\email
{schild@physik.uni-bielefeld.de}
\affiliation{Fakult\"at f\"ur Physik, Universit\"at Bielefeld,
Universit\"atsstra{\ss}e 25, D-33615 Bielefeld, Germany\\
and\\
Max-Planck-Institut f\"ur Physik (Werner-Heisenberg-Institut),
F\"ohringer Ring 6, D-80805 M\"unchen, Germany}

\begin{abstract}
We investigate the mass range of the quark-antiquark fluctuations of
the photon that are active in producing the total photoabsorption
cross section in the color dipole picture, emphasizing the notions
of color transparency and saturation. We consider the implications
of measurements at future extensions of the available 
electron-proton-scattering energy.
\end{abstract}
\maketitle

\section{Introduction}
\renewcommand{\theequation}{\arabic{section}.\arabic{equation}}
\setcounter{equation}{0}

Deep inelastic scattering (DIS) of electrons on protons at
low values of the Bjorken variable $x \equiv x_{bj}\cong Q^2/W^2 \lsim 0.1$
(where $Q^2$ refers to the photon virtuality and $W$ to the
photon-proton center-of-mass energy)
is a two-step process: transition, or fluctuation in modern jargon,
of the photon into on-shell quark-antiquark $(q \bar q)$ states,
$\gamma^* \to q \bar q$, of mass $M_{q \bar q}$, and subsequent
scattering of these states on the proton. In terms of the photon-proton
(virtual) forward Compton scattering amplitude, the $q \bar q$ states
interact with the proton via (color) gauge-invariant two-gluon exchange:
the color dipole picture (CDP)\footnote{Compare refs. \cite{PRD85, IJMPA31} for an
extensive list of references}. A model-independent 
analysis\footnote{``Model-independent'' means that the results for the
photoabsorption cross section do not depend on a parameter-dependent explicit
ansatz for the $q \bar q$-dipole-proton interaction, except for a decent
unitarity-preserving high-energy behavior.} \cite{PRD85} shows that
the photoabsorption cross section, $\sigma_{\gamma^*p} (W^2,Q^2)$, depends
on the low-x scaling variable $\eta (W^2,Q^2) = (Q^2+m^2_0)/\Lambda^2_{sat} (W^2)$
\cite{Surrow}
via $\sigma_{\gamma^*p} (W^2,Q^2) = \sigma_{\gamma^*p} (\eta (W^2,Q^2)) \sim
1/\eta(W^2,Q^2)$ for $\eta (W^2,Q^2) \gsim 1$ (``color transparency''),
while $\sigma_{\gamma^*p} (W^2,Q^2) = \sigma_{\gamma^*p} (\eta (W^2,Q^2)) \sim
\ln (1/\eta (W^2,Q^2))$ (``saturation'') for $\eta (W^2,Q^2) \lsim 1$ \footnote{The behavior
in terms of $1/\eta (W^2,Q^2)$ is valid except for a logarithmic, $\ln W^2$, energy
dependence of the dipole cross section
$\sigma_{(q \bar q)p} (W^2)$.   See the discussion on the relation of
the dipole cross section to $(Q^2 = 0)$ photoproduction to be given in Section IV.}.
The ``saturation scale'' $\Lambda^2_{sat} (W^2)$ increases with a small power
of $W^2$, and $m_0$ is a constant mass, in the case of light quarks
somewhat below the $\rho^0$-meson mass. Any specific parameter-dependent
ansatz \cite{PRD85, IJMPA31, Surrow} for the $q \bar q$-dipole-proton cross
section has to interpolate between the $1/\eta (W^2,Q^2)$ and the 
$\ln (1/\eta (W^2,Q^2))$ dependence.

The validity of the CDP rests on the condition that in the $\gamma^* \to 
q \bar q$ transition the proton-rest-frame energy imbalance $\Delta E$ between the photon of
virtuality $q^2 \equiv -Q^2 \le 0$ and the $q \bar q$ state of invariant mass 
squared   $M^2_{q \bar q} > 0$ be  small for sufficiently large 
 $W^2\gg M_p^2, Q^2$,
\be
\Delta E \simeq \frac{Q^2+M^2_{q \bar q}}{W^2} M_p \ll M_p,
\label{eq:1.1}
\ee
or
\be
\frac{Q^2+M^2_{q \bar q}}{W^2} \ll 1.
\label{eq:1.2}
\ee
Compare Appendix A.  The condition (\ref{eq:1.2})
\begin{itemize}
\item[i)] restricts the kinematical range of the CDP to
$x \cong Q^2/W^2 \ll 1$, and it
\item[ii)] contains the dynamical restriction  of $M^2_{q \bar q}/W^2
\ll 1$ from  generalized vector dominance (GVD).
The transition of the photon, $\gamma^* \to q \bar q$, to
a finite range of masses, $M_{q \bar q}$, saturates the $\gamma^*$-proton
cross section for given photon virtuality $Q^2$ and energy $W$ with
$x \cong Q^2/W^2 \lsim 0.1$.
\end{itemize}

\begin{figure}
\includegraphics[width=8cm]{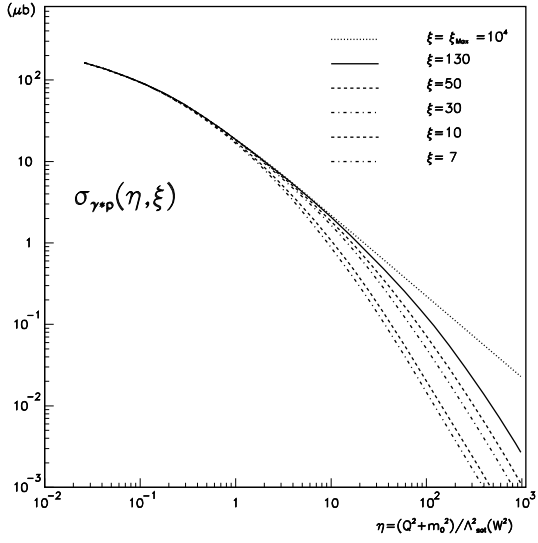}
\caption{The theoretical results for the photoabsorption cross section
$\sigma_{\gamma^*p} (\eta (W^2,Q^2), \xi)$ in the CDP as a function of
the low-x scaling variable $\eta (W^2,Q^2) = (Q^2 + m^2_0)/\Lambda^2_{sat}
(W^2)$ for different values of the parameter $\xi$ that determines the
(squared) mass range $M^2_{q \bar q} \le m^2_1 (W^2) = \xi \Lambda^2_{sat}
(W^2)$ of the $\gamma^* \to q \bar q$ fluctuations that are taken into
account. The experimental results\cite{HERA}   for 
 $\sigma_{\gamma^*p}(\eta(W^2,Q^2),\xi)$ 
 lie on the full line corresponding
to $\xi = \xi_0 = 130$, compare refs. \cite{PRD85}, \cite{IJMPA31}.}
\label{fig:xi}
\end{figure}

It is the purpose of this paper to present a detailed investigation
of the mass range of $\gamma^* \to q \bar q$ fluctuations responsible for,
or actively producing, the photoabsorption cross section at different
values of the kinematic variables $W^2, Q^2$ and $\eta (W^2,Q^2)$.
We emphasize the different regions of $\eta (W^2,Q^2)$ related to
color transparency and saturation. We comment on the impact of a future
extension of the ep energy range, and on the determination  of the asymptotic
energy dependence of $(Q^2=0)$ photoproduction from  the measured
$W$-dependence of the dipole cross section.

\section{The Photoabsorption Cross Section}
\renewcommand{\theequation}{\arabic{section}.\arabic{equation}}
\setcounter{equation}{0}

We start by a discussion of the results for the photoabsorption cross
section, $\sigma_{\gamma^*p} (W^2,Q^2)$, in the CDP that are shown
in Fig. 1, reproduced from ref. \cite{IJMPA31}. The results are obtained
from the explicit analytic expression for $\sigma_{\gamma^*p} (W^2,Q^2)$,
\footnote{To indicate the dependence of  $\sigma_{\gamma^*p} (W^2,Q^2)$
on $\xi$, we frequently use, as in Fig.1, the notation
$\sigma_{\gamma^*p} (W^2,Q^2) \equiv \sigma_{\gamma^*p} (W^2,Q^2,\xi)$
as well as  $\sigma_{\gamma^*p} (\eta(W^2,Q^2),\xi)$.  The dependence on $\xi$
is contained in $G_{T,L}(u\equiv \xi/\eta(W^2,Q^2))$, see (\ref{eq:2.9}) and
(\ref{eq:2.10}) below.}\cite{IJMPA31}
\bqa
\sigma_{\gamma^*p} (W^2,Q^2) & = & \sigma_{\gamma^*_Lp} (W^2,Q^2) +
\sigma_{\gamma^*_Tp} (W^2,Q^2)  \nonumber \\
& = & \frac{\alpha R_{e^+e^-}}{3 \pi} 
\sigma^{(\infty)} (W^2)  \nonumber \\
&&\times\left( I^{(1)}_T \left( \frac{\eta}{\rho}, \frac{\mu}{\rho} \right)
G_T (u) \right. \nonumber \\
&&~~~~~\left.+ I^{(1)}_L
(\eta, \mu) G_L (u)
\right),\nonumber\\
&&
\label{eq:2.1}
\eqa
derived from an ansatz for the W-dependent dipole cross 
section\footnote{Following the suggestion of the (anonymous) referee, 
in  Appendix B, we present a brief (critical) discussion
on the approach of ``geometric scaling'' based on an $x \simeq Q^2/W^2$-dependent, 
and accordingly $Q^2$-dependent, ansatz for the dipole cross section.} 
that essentially, via
coupling of the quark-antiquark state to two gluons, comprises the
color-gauge-invariant interaction of the $q \bar q$ dipole with the
gluon-field in the nucleon. 
In (\ref{eq:2.1}), $R_{e^+e^-}=3\sum_qQ_q^2$, where $q$ runs over the active quark
flavors, and $Q_q$ denotes the quark charge.
The smooth transition to $Q^2 = 0$
photoproduction in (\ref{eq:2.1}) allows one\cite{IJMPA31}  to replace 
$\sigma^{(\infty)} (W^2)$, which stems from the normalization of the
dipole cross section, by the photoproduction cross section, i.e.
\bqa
\sigma_{\gamma^*p} (W^2, Q^2) && = \frac{\sigma_{\gamma p} (W^2)}
{\lim\limits_{\eta \to
\mu (W^2)} I_T^{(1)} 
\left( \frac{\eta}{\rho},
\frac{\mu (W^2)}{\rho}\right)}  \nonumber \\
&& \hspace*{-1.4cm}\times\left( I^{(1)}_T  \left(
\frac{\eta}{\rho}, \frac{\mu}{\rho} \right) G_T (u) + I^{(1)}_L (\eta, \mu)
G_L (u) \right). ~~
\label{eq:2.2}
\eqa
We note that $I_L^{(1)} (\eta, \mu)$ vanishes in the photoproduction
$Q^2 = 0$ limit of $\eta(W^2,Q^2=0)=m_0^2/\Lambda^2_{sat}(W^2) \equiv \mu(W^2)$,
and  $G_T (u\equiv{\xi\over\eta}) \simeq 1$,
and for later reference we also note
\be
\lim_{\eta \to \mu (W^2)} I^{(1)}_T \left( \frac{\eta}{\rho},
\frac{\mu (W^2)}{\rho} \right) = \ln \frac{\rho}{\mu (W^2)}.
\label{eq:2.3}
\ee
The general explicit analytic expressions for the functions $I^{(1)}_T
\left( \frac{\eta}{\rho},\frac{\mu (W^2)}{\rho} \right)$ and
$I^{(1)}_L (\eta, \mu)$ are not needed for the ensuing discussions,
and we refer to ref. \cite{IJMPA31}, while $G_T \left( u\equiv\frac{\xi}{\eta}
\right)$ and $G_L \left( u\equiv\frac{\xi}{\eta} \right)$ will be given
in (\ref{eq:2.9}) and (\ref{eq:2.10}) below. 
The numerical results for the photoabsorption cross section
in Fig. 1 are obtained by numerical evaluation of (\ref{eq:2.2})
upon insertion of a $\left( \ln (W^2) \right)^2$ fit to the experimental results for
the photoproduction cross section $\sigma_{\gamma p} (W^2)$ from
the Particle Data Group\cite{PDG}. The results in Fig. \ref{fig:xi} were
obtained for $W = 275 {\rm GeV}$.   Compare Section IV, and Fig.
\ref{fig:eta} in section IV, for the (weak) W dependence  of 
$\sigma_{\gamma^*p} (\eta(W^2,Q^2), W^2)$ due to $\sigma^{(\infty)}(W^2)$ in (\ref{eq:2.1})
and  $\sigma_{\gamma p} (W^2)$ in (\ref{eq:2.2}).

Before going into more detail, we note
that the full curve in Fig. 1, which  for the parameter $\xi$ corresponds 
to the choice of
$\xi = \xi_0 = 130$,  is consistent with and provides
a representation of the full set of experimental data on 
$\sigma_{\gamma^*p} (W^2,Q^2)$, compare Fig. 9 in ref. \cite{PRD85}.

In (\ref{eq:2.1}) and (\ref{eq:2.2}), the low-x scaling variable $\eta (W^2,Q^2)$ is
given by
\be
\eta \equiv \eta (W^2,Q^2) = \frac{Q^2 + m^2_0}{\Lambda^2_{sat} (W^2)},
\label{eq:2.4}
\ee
with
\be
\mu \equiv \mu(W^2) = \eta (W^2,Q^2 = 0) = \frac{m^2_0}{\Lambda^2_{sat}
(W^2)},
\label{eq:2.5}
\ee
the saturation scale, $\Lambda^2_{sat} (W^2)$, being parametrized by
\be
\Lambda^2_{sat} (W^2) = C_1 \left( \frac{W^2}{1 {\rm GeV}^2} \right)^{C_2},
\label{eq:2.6}
\ee
and numerically, we have
\bqa
m^2_0 & = & 0.15 {\rm GeV}^2, \nonumber \\
C_1 & = & 0.31;~~~C_2 = 0.27.
\label{eq:2.7}
\eqa
The parameter $\rho$ is related to the longitudinal-to-transverse
ratio $R (W^2,Q^2)$ of the photoabsorption cross section, and
approximately we have $R(W^2,Q^2) \simeq 1/2 \rho$ for
$\eta (W^2,Q^2) \gg \mu (W^2)$, while $R(W^2,Q^2) = 0$ for $Q^2 = 0$.
The total cross section is fairly insensitive to the value of $\rho$, and  
the evaluation presented in Fig. 1 is based  \cite{IJMPA31} on 
$\rho = {4\over 3}$.

Our main concern in the rest of this Section and the following one will
center around the dependence of the cross section (\ref{eq:2.2}) on
the constant parameter $\xi$ that, by definition, restricts the masses
of the contributing $q \bar q$ states via

\be
M^2_{q \bar q} \le m^2_1 (W^2) = \xi \Lambda^2_{sat} (W^2).
\label{eq:2.8}
\ee
The dependence on $\xi$ in (\ref{eq:2.1}) and (\ref{eq:2.2}) is contained\cite{IJMPA31} in the
functions $G_{T,L} (u \equiv \xi/\eta(W^2,Q^2))$,
\be
G_T (u) =  \frac{2u^3 + 3 u^2 + 3u}{2 (1+u)^3} \simeq
\left\{ \begin{array}{l@{\quad,\quad}l}
\frac{3}{2} \frac{\xi}{\eta} & (\eta \gg \xi), \\
1 - \frac{3}{2} \frac{\eta}{\xi} & (\eta \ll \xi),
\end{array} \right.
\label{eq:2.9}
\ee
and
\be
G_L (u) = \frac{2u^3 + 6u^2}{2 (1+u)^3} \simeq
\left\{ \begin{array}{l@{\quad,\quad}l}
3 \left( \frac{\xi}{\eta}\right)^2 & (\eta \gg \xi),\\
1 - 3 \left( \frac{\eta}{\xi} \right)^2 & (\eta \ll \xi).
\end{array} \right.
\label{eq:2.10}
\ee

We turn to a more detailed qualitative discussion of the theoretical predictions in Fig. 1.

The parameter$\xi$ is bounded by $\xi\le \xi_{Max}(W^2)$, where $\xi_{Max}(W^2)$ 
corresponds to the upper limit of $m_1^2(W^2)\cong W^2$
in (\ref{eq:2.8});  the contributing $q\bar q$-dipole masses cannot exceed the total 
available $(q\bar q)p$ center-of-mass energy $W$.  Accordingly, we have
\bq
    \xi_{Max}=W^2/\Lambda^2_{sat}(W^2),
\label{eq:2.11}
\eq
as well as 
\be
   \frac{\eta(W^2,Q^2)}{\xi_{Max}(W^2)}\cong \frac{Q^2}{W^2}\cong x_{bj},
   ~~~~({\rm for}~~Q^2\gg m_0^2),
\label{eq.2.12}
\ee
where $x_{bj}\lsim 0.1$, and 
\bqa
   &&G_{T,L}(\xi_{Max}(W^2)/\eta(W^2,Q^2)) \nonumber \\
   &\cong&   G_{T,L}(\xi_{Max}(W^2)/\eta(W^2,Q^2) \to \infty).
\label{eq:2.13}
\eqa
The total photoabsorption cross section (\ref{eq:2.2})  for $\xi=\xi_{Max}(W^2)$ becomes
\bq
   \sigma_{\gamma^*p} (W^2,Q^2,\xi=\xi_{max})\cong  
   \sigma_{\gamma^*p}(\eta(W^2,Q^2), \xi \to\infty).
\label{eq:2.14}
\eq
Specificaly, in Fig.1, we have $W=275$ GeV and 
$\xi_{Max}\simeq 10^4\gg\eta(W^2,Q^2)$
implying the validity of (\ref{eq:2.14}).

For $\xi=\xi_{Max}$, from (\ref{eq:2.8}) with (\ref{eq:2.11}), the upper bound on 
$q\bar q$-dipole masses becomes
\be
     \frac{M^2_{q \bar q}}{W^2} \le 1.
\label{eq:2.15}
\ee  
The prediction for the photoabsorption cross section in Fig.1 for $\xi=\xi_{Max}=10^4 $
includes contributions from $q\bar q$ masses that strongly violate the fundamental 
condition on $\Delta E/M_p\ll 1$ in (\ref{eq:1.2}).

Turning to $\xi=\xi_0=130\ll \xi_{Max}$, in distinction from (\ref{eq:2.15}), we find
\be
\frac{M^2_{q \bar q}}{W^2} \le \xi_0 \frac{\Lambda^2_{sat} (W^2)}{W^2}
\simeq 0.01,
\label{eq:2.16}
\ee
where $W=275$ GeV from Fig.1 was inserted.
The mass range of contributing $q\bar q$ states is consistent with $\Delta E/M_p\ll 1$.

The experimental results on the photoabsorption cross section agree with 
the theoretical prediction for $\xi=\xi_0=130$ in Fig.1.  The distinctive 
difference between the theoretical cross section for $\xi=\xi_{Max}$ and 
the experimentally verified one for $\xi=\xi_0=130$ seen for $\eta(W^2,Q^2)\gsim 10$ 
in Fig.1, explicitly demonstrates that the $(q\bar q)p$ interaction is due to 
$q \bar q$-dipole states that are limited in mass by $M_{q\bar q}^2\le \xi_0
 \Lambda^2_{sat}(W^2)$.  The experimental data on $\sigma_{\gamma^*p}(W^2,Q^2)$
confirm the validity of the energy imbalance for $\gamma^*\to q \bar q$ transition 
in (\ref{eq:1.2})  

We turn to the theoretical results for $\xi<\xi_0$  also shown in Fig.1.
From the difference of the cross sections for $\xi<\xi_0$ and $\xi=\xi_0=130$ at
$\eta(W^2,Q^2)\gsim 1$, we conclude that high-mass $q\bar q$-dipole
contributions are definitely necessary to saturate the forward-Compton-scattering
amplitude.   

The theoretical predictions for 
$\sigma_{\gamma^*p}(W^2,Q^2,\xi)$ for $\xi<\xi_0$ with decreasing $\eta(W^2,Q^2)$,
however, 
show a tendency to converge towards the results obtained for $\xi=\xi_0$.  
This behavior indicates that with decreasing $\eta(W^2,Q^2)$ 
( or decreasing $Q^2$  at fixed $W^2$), nevertheless,  only $q\bar q$ states with decreasing 
mass squared $M_{q\bar q}^2\le \xi \Lambda^2_{sat}(W^2)< \xi_0\Lambda^2_{sat}$ are
actually relevant, or "active" , for producing the total photoabsorption
cross section.  

A detailed investigation of the mass range of active 
$\gamma^*\to q\bar q$  transitions will be the subject of Section III.

\section{The mass range of active $\mathbf{\gamma^*\to q \bar q}$ fluctuations}
\renewcommand{\theequation}{\arabic{section}.\arabic{equation}}
\setcounter{equation}{0}

We turn to quantifying the mass range of those $q \bar q$ states that are
responsible for the major part of the experimentally observed cross
section $\sigma_{\gamma^*p} (W^2,Q^2)$ in Fig. 1. The range of 
contributing $q \bar q$ masses $m^2_0  \le M^2_{q\bar q}\le m^2_1 (W^2) \le \xi
\Lambda^2_{sat} (W^2)$ being determined by the parameter $\xi$,
we search for the value of $\xi$ that yields a (substaintial) fraction of
$1 - \epsilon$, where $\epsilon = const. \ll 1$, of the
photoabsorption cross section $\sigma_{\gamma^*p} (W^2,Q^2)$.

Employing the expression for $\sigma_{\gamma^*p} (W^2,Q^2)$ in
(\ref{eq:2.2}) together with the approximate expressions for
$G_{T,L} (u \equiv \xi/\eta (W^2,Q^2))$ in (\ref{eq:2.9}) and
(\ref{eq:2.10}), we find that the dependence of $\sigma_{\gamma^*p}
(W^2,Q^2)$ on $\eta (W^2,Q^2)/\xi$ for $\eta (W^2,Q^2)/\xi  \ll 1$
is approximately given by the factor $1 - (3/2) \eta
(W^2,Q^2) / \xi$ in (\ref{eq:2.9}), i.e. upon employing (\ref{eq:2.14}), 
\bqa
\sigma_{\gamma^*p} (W^2,Q^2,\xi)& = & \sigma_{\gamma^*p}
(\eta(W^2,Q^2), \xi_{Max})  \nonumber \\
&& \times  \left(1 - \frac{3}{2} \frac{\eta (W^2,Q^2)}{\xi}
\right).
\label{eq:3.1}
\eqa
The experimentally observed cross section for $\eta (W^2,Q^2)/\xi_0  
\ll 1$  is represented by evaluating (\ref{eq:3.1}) for
$\xi = \xi_0 = 130$,
\bqa
\sigma_{\gamma^*p} (W^2,Q^2,\xi_0) & = & \sigma_{\gamma^*p}
(\eta(W^2,Q^2) , \xi_{Max})  \nonumber \\
&& \times  \left( 1 - \frac{3}{2}
\frac{\eta (W^2,Q^2)}{\xi_0} \right).
\label{eq:3.2}
\eqa
A fraction of $1 - \epsilon$ of the experimentally observed 
cross section (\ref{eq:3.2}) accordingly is associated with
a value of $\xi$ such that $\sigma_{\gamma^*p} (W^2,Q^2,\xi)$
deviates from  $\sigma_{\gamma^*p} (W^2,Q^2,\xi_0)$ by the
factor $(1- \epsilon)$,
\be
\sigma_{\gamma^*p} (W^2,Q^2, \xi) = \sigma_{\gamma^*p} (W^2,
Q^2, \xi_0) (1 - \epsilon).
\label{eq:3.3}
\ee
Substitution of (\ref{eq:3.1}) and (\ref{eq:3.2}) into (\ref{eq:3.3})
yields
\be
1 - \frac{3}{2} \frac{\eta (W^2,Q^2)}{\xi} = \left( 1 - \frac{3}{2}
\frac{\eta (W^2,Q^2)}{\xi_0} \right) (1 - \epsilon)
\label{eq:3.4}
\ee
or
\be
\xi = \frac{3}{2 \epsilon} \eta (W^2,Q^2) \frac{1}{1 +
\frac{3 \eta (W^2,Q^2)}{2 \epsilon \xi_0}(1-\epsilon)}.
\label{eq:3.5}
\ee
This is the value of  $\xi$ that, according to (\ref{eq:3.3}),  for given $\eta (W^2,Q^2)
\ll \xi_0$ yields a fraction of $1 - \epsilon$ of the photoabsorption
cross section $\sigma_{\gamma^*p} (W^2,Q^2, \xi_0)$. For
$\epsilon \to 0$, consistently, we have $\xi \to \xi_0$ in (\ref{eq:3.5}), or
$m^2_1 (W^2) = \xi_0~\Lambda^2_{sat} (W^2)$, corresponding to
the experimentally observed cross section. 

For $\eta (W^2,Q^2)/\xi_0
\ll \epsilon$, we may approximate (\ref{eq:3.5}) by
\be
\xi = \frac{3}{2 \epsilon} \eta (W^2,Q^2),
\label{eq:3.6}
\ee
and this approximation will be adopted subsequently. For e.g.
$\epsilon = 0.1$, from (\ref{eq:3.6}), we have $\xi = 15 \eta
(W^2,Q^2)$. For any given $\eta (W^2,Q^2) \ll \xi_0$, from (\ref{eq:3.5})
or (\ref{eq:3.6}), we obtain a value of $\xi$ that for e.g.
$\epsilon = 0.1$ provides 90 \% of the experimentally verified 
photoabsorption cross section.

In terms of $m^2_1 (W^2) = \xi \Lambda^2_{sat} (W^2)$, from
(\ref{eq:3.6}), we have
\bqa
m^2_0 \le M^2_{q\bar q} \le m^2_1 &&= \frac{3}{2 \epsilon} \eta
(W^2,Q^2) \Lambda^2_{sat} (W^2) \nonumber \\
&&= \frac{3}{2 \epsilon} (Q^2 + m^2_0).
\label{eq:4}
\eqa
For any $W^2$ and $Q^2$ with $\eta(W^2,Q^2) \ll \xi_0$, the
constraint (\ref{eq:4}) determines the mass range of $q \bar q$
dipole states that are essential for the cross section in the sense of
providing a fraction of magnitude $1 - \epsilon$ of the
photoabsorption cross section $\sigma_{\gamma^*p} (W^2,Q^2)$.
In other words, the dominant contribution to the photoabsorption
cross section for fixed $\eta (W^2,Q^2) \ll \xi_0$ is due to
$q \bar q$ states that have masses below the limit given in
(\ref{eq:4}). The masses of these ``active'' $q \bar q$ states are
restricted by the value of the photon virtuality $Q^2$
according to (\ref{eq:4}). A fixed value of $Q^2$ is uniquely
associated with a fixed $q \bar q$-dipole-mass range.

In Table \ref{tab:tab1}, for the choice of $\epsilon = 0.1$, we show the
results of a numerical evaluation of the upper limit $m^2_1$
from (\ref{eq:4}) for various values of $\eta (W^2,Q^2) \ll \xi_0$ 
and for energies in the range of $W \lsim 300$ GeV explored
at HERA \cite{HERA}, and at the energy $W = 10^4$ GeV recently discussed 
in view of future collider projects \cite{Caldwell}.
For the saturation scale $\Lambda^2_{sat} (W^2)$, and for 
$m^2_0$, we use the parameters adjusted to the experimental data from
HERA for $x_{bj} \cong Q^2/W^2 \lsim 0.1$, compare (\ref{eq:2.7}).
\begingroup
\squeezetable
\begin{table}
\caption{\label{tab:tab1}The $(\eta, W)$ matrix elements give the numerical 
values from (\ref{eq:4}) 
with $\epsilon = 0.1$ of the mass range
$m_0 \le M_{qq} < m_1$ of $\gamma^* \to q \bar q$ transitions
for fixed values of $\eta (W^2,Q^2) = (Q^2 + m^2_0)/\Lambda^2_{sat}
(W^2)$ and energy $W$. At fixed $\eta$, with increasing
energy $W$, increasing $q \bar q$ masses determine the cross section.
At fixed $W$, with decreasing $\eta (W^2,Q^2)$ smaller masses
determine the cross section.}
\begin{ruledtabular}
\begin{tabular}{llll}
W [Gev] & 30 & 300 &$10^4$ \\
$\Lambda^2_{sat} (W^2) [{\rm GeV}^2]$ & 1.95 &6.75 & 44.8 \\
$\eta_{Min} (W^2)$ & $7.6 \times 10^{-2}$ & $2.2 \times 10^{-2}$
& $3.3 \times 10^{-3}$\\ 
\hline
&&&\\
 & $Q^2 = 1.8~ {\rm GeV}^2$ & $ Q^2 = 6.9~ {\rm GeV}^2$
& $Q^2 = 44.7~ {\rm GeV}^2$ \\
$\eta = 1$ & $m^2_1 = 29~ {\rm GeV}^2$ &$m^2_1 = 101~ {\rm GeV}^2$
& $m^2_1 = 672~ {\rm GeV}^2$ \\
 & $m_1 = 5.4~ {\rm GeV}$ & $m_1 = 10~ {\rm GeV}$ & $m_1 = 25~ {\rm GeV}$\\
&&&\\
 & $Q^2 = 4.5 \times 10^{-2}$ & $ Q^2 = 0.53~ {\rm GeV}^2$
& $Q^2 = 4.3~ {\rm GeV}^2$ \\
$\eta = 0.1$ & $m^2_1 = 2.9~ {\rm GeV}^2$ &$m^2_1 = 10.1~ {\rm GeV}^2$
& $m^2_1 = 67~ {\rm GeV}^2$ \\
 & $m_1 = 1.7~ {\rm GeV}$ & $m_1 = 3.2~ {\rm GeV}$ & $m_1 = 8.2~ {\rm GeV}$\\
&&&\\
& \multicolumn{3}{c}{$Q^2 = 0$}\\
$\eta = \eta_{Min}$ &\multicolumn{3}{c}{$m^2_1 = 2.25~ {\rm GeV}^2$} \\
& \multicolumn{3}{c}{$m_1 = 1.5~ {\rm GeV}$}\\
\end{tabular}
\end{ruledtabular}
\end{table}
\endgroup

According to Table \ref{tab:tab1}, for various fixed values of $\eta (W^2,Q^2)$, with
increasing $W$, we find the expected increase of the upper limit of the masses of
relevant $q \bar q$ states, $M^2_{q \bar q} \lsim m^2_1$. For 
e.g. $\eta (W^2,Q^2) = 1$, we have $M^2_{q \bar q} \le m^2_1 = 29
~{\rm GeV}^2$ at $W = 30$ GeV, and $M^2_{q \bar q} \le m^2_1 = 101~{\rm GeV}^2$ at
$W = 300~{\rm GeV}$, and finally $M^2_{q \bar q} \le m^2_1 = 672~{\rm GeV}^2$
at $W = 10^4~{\rm GeV}$. With decreasing $\eta (W^2,Q^2)$ at fixed $W$, the 
decrease in $Q^2$ is accompanied by a decrease in $m^2_1$, leading to
$M^2_{q \bar q} \le m^2_1 = 2.25~{\rm GeV}^2$ at 
$\eta (W^2,Q^2) = \eta (W^2, Q^2=0) \equiv \eta_{Min}$. It is
amusing to note that the value of $m_1 = 1.5~{\rm GeV}$ practically
coincides with the value of $m_1 = 1.4~{\rm GeV}$ from the 1972
Generalized Vector Dominance (GVD) interpretation \cite{Sakurai, APP}
of the first data on DIS from the SLAC-MIT collaboration \cite{SLAC-MIT}.

\begin{table}
\caption{\label{tab:tab2}The $(Q^2,W)$ matrix elements are the values of
$\eta (W^2,Q^2) = (Q^2 + m^2_0)/\Lambda^2_{sat} (W^2)$.
A fixed value of $Q^2$ is associated with a fixed (squared) mass
range, $m^2_0 \le M^2_{q \bar q} \le m^2_1$. With increasing energy
$W$, for fixed $Q^2$ and fixed $M^2_{q \bar q} < m^2_1$, the transition from 
color transparency $(\eta \gg 1)$ to saturation $(\eta \ll 1)$
takes place. For $Q^2 \simeq 0$, hadronlike saturation behavior
occurs for all values of $W$ shown. With decreasing $Q^2$ at
fixed $W$ decreasing masses, $M^2_{q \bar q} < m^2_1$ determine
the cross section.}
\begin{ruledtabular}
\begin{tabular}{llll}
W [Gev] & 30 & 300 &$10^4$ \\
$\Lambda^2_{sat} (W^2) [{\rm GeV}^2]$ & 1.95 &6.75 & 44.8 \\
\hline
&&&\\
$Q^2 = 10~ {\rm GeV}^2$ & 5.2 & 1.5 & $2.3 \times 10^{-1}$ \\
$m^2_1 = 152~ {\rm GeV}^2$ & & & \\
$m_1 = 12.3~ {\rm GeV}$ & & & \\
&&&\\
$Q^2 = 2~ {\rm GeV}^2$ & 1.1 & $3.2 \times 10^{-1}$ & $4.8 \times 10^{-2}$ \\
$m^2_1 = 32.3~ {\rm GeV}^2$ & & & \\
$m_1 = 5.68~ {\rm GeV}$ & & & \\
&&&\\
$Q^2 = 0 $ & $7.7 \times 10^{-2}$ & $2.2 \times 10^{-2}$ 
& $3.3 \times 10^{-3}$ \\
$m^2_1 = 2.25~ {\rm GeV}^2$ & & & \\
$m_1 = 1.5~ {\rm GeV}$ & & & \\
\end{tabular}
\end{ruledtabular}
\end{table}

In Table \ref{tab:tab2}, we present the values of the scaling variable
$\eta (W^2,Q^2)$ corresponding to fixed values of $Q^2$ (and of
$m^2_1$ according to (\ref{eq:4}) with $\epsilon = 0.1$), 
for different values of $W$ chosen as in Table \ref{tab:tab1}.
The Table illustrates that an identical fixed mass range,
defined by $m^2_0 \le M^2_{q \bar q} \le m^2_1$, is responsible
for cross sections in the color transparency region and the 
saturation region; e.g. for $Q^2 = 2~{\rm GeV}^2$ and $m_1 =
5.68~{\rm GeV}$, we see the transition from $\eta = 1.1 \gsim 1$
at $W = 30~{\rm GeV}$ to $\eta = 4.8 \times 10^{-2} \ll 1$ that
is reached at $W = 10^4~{\rm GeV}$. As a consequence of the 
two-gluon color-dipole interaction, a massive $q \bar q$
state of mass $ m_0\le M_{q \bar q}\le m_1$, dependent on the energy $W$, 
either interacts with a small cross
section (color transparency), $\sigma_{\gamma^*p} (W,Q^2) \sim 
1/\eta (W^2, Q^2)$, or with a moderately large one (saturation), 
$\sigma_{\gamma^*p} (W^2,Q^2) \sim \ln (1/\eta (W^2,Q^2))$.

\section{The Extraction of the $Q^2 = 0$ Photoproduction Cross Section}
\renewcommand{\theequation}{\arabic{section}.\arabic{equation}}
\setcounter{equation}{0}

So far in this paper, we were concerned with the $\eta (W^2,Q^2)$
dependence of the photoabsorption cross section and its connection
with the mass range of contributing $\gamma^* \to q \bar q$ transitions.
As previously mentioned, and explicitly seen in (\ref{eq:2.1}) and
(\ref{eq:2.2}), there is a deviation from a pure $\eta (W^2,Q^2)$
dependence that originates from the $W^2$ dependence of the dipole
cross section. We recall that the results in (\ref{eq:2.1}) and
(\ref{eq:2.2}) follow by specializing\cite{PRD85, IJMPA31}   
the generic two-gluon-exchange form of the dipole cross section
\bqa
\sigma_{(q \bar q)} (\vec r_\bot, z (1-z), W^2) = && \int d^2 \l_\bot
\tilde \sigma \left( \vec l^{~2}_\bot, z(1-z),W^2 \right) \nonumber \\
&& \times \left( 1 - e^{-\vec l_\bot \cdot \vec r_\bot} \right)
\label{eq:4.1}
\eqa
via the ansatz
\bqa
\tilde \sigma \left( \vec l^{~2}_\bot, z (1-z), W^2 \right) &=&
\frac{\sigma^{(\infty)} (W^2)}{\pi}   \nonumber \\
&& \hspace*{-1.0cm}\times\delta \left( \vec l^{~2}_\bot
-z (1-z) \Lambda^2_{sat} (W^2) \right).
\label{eq:4.2}
\eqa
The connection between the normalization of the dipole cross section,
$\sigma^{(\infty)} (W^2)$, which coincides with the limit of the 
dipole cross section for $\Lambda^2_{sat} (W^2) \vec r^{~2}_\bot
\to \infty$, and the $Q^2 = 0$ photoproduction cross section, 
$\sigma_{\gamma p} (W^2)$, is implicitly contained
in (\ref{eq:2.1}) to (\ref{eq:2.3}), i.e.
\be
\sigma^{(\infty)} (W^2) = \frac{3\pi}{\alpha R_{e^+e^-}}
\frac{1}{\ln \frac{\Lambda^2_{sat} (W^2)}{m^2_0}}
\sigma_{\gamma p} (W^2),
\label{eq:4.3}
\ee
or
\be
\sigma_{\gamma p} (W^2) = \frac{\alpha R_{e^=e^-}}{3 \pi} 
\sigma^{(\infty)} (W^2) \ln \frac{\Lambda^2_{sat} (W^2)}{m^2_0},
\label{eq:4.4}
\ee
where we put $\rho=1$ for simplicity.
According to (\ref{eq:4.3}) and (\ref{eq:4.4}), the dipole cross section
$\sigma^{(\infty)} (W^2)$ and the photoproduction cross section are
uniquely related to each other. For
\begin{itemize}
\item[i)] $\sigma^{(\infty)} (W^2) =  const.$, from (\ref{eq:4.4}) and (\ref{eq:2.2})
with (\ref{eq:2.3}),  we have strict validity of scaling in $\eta (W^2,Q^2)$,
i.e. $\sigma_{\gamma^*p} (W^2,Q^2) = \sigma_{\gamma^*p} (\eta(W^2,Q^2))$,
and $\sigma_{\gamma p} (W^2) \sim \ln W^2$, while for
\item[ii)] $\sigma^{(\infty)} (W^2) \sim \ln W^2$, we have logarithmic
violation of scaling in $\eta (W^2,Q^2)$ for $\sigma_{\gamma^*p} (W^2,Q^2)$,
while $\sigma_{\gamma p} (W^2) \sim (\ln W^2)^2$, and finally,
\item[iii)] a "hadronlike" dipole cross section, $\sigma^{(\infty)} (W^2) \sim (\ln W^2)^2$,
 leads to $\sigma_{\gamma p} (W^2) \sim (\ln W^2)^3$.  From a different
angle, a potential dependence as $(\ln W^2)^3$ was recently considered by
Mueller\cite{mueller}.
\end{itemize}
\begin{figure}
\includegraphics[width=8cm]{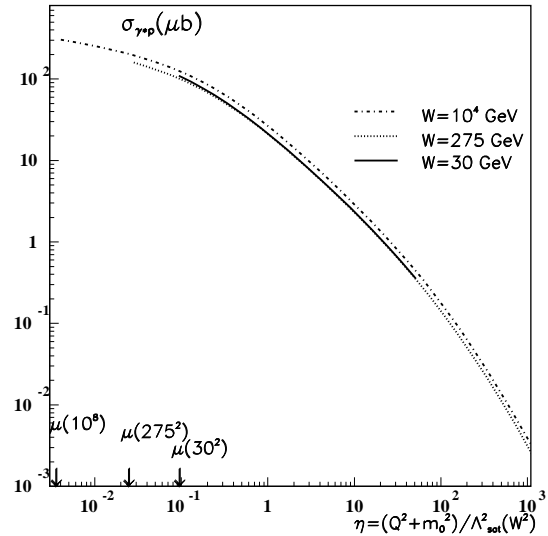}
\caption{The theoretical results for the photoabsorption cross section
as a function of $\eta (W^2,Q^2)$ for different values of $W$. The dependence
on $W$ is due to logarithmic $\eta$-scaling violations  $\sigma^{(\infty)}
(W^2) \sim \ln W^2$, compare case  ii) in the main text. For $\xi$ the value $\xi = \xi_0 = 130$ is used,
and the results for $W = 275$ GeV are identical to the results shown by the
full line in Fig. \ref{fig:xi}. By definition, $\mu (W^2) \equiv m^2_0/\Lambda^2_{sat}
(W^2)$.}
\label{fig:eta}
\end{figure}

In Fig. \ref{fig:eta}, we show the results corresponding to case ii), based
on the high-energy extrapolation in $W$ of the fit to photoproduction
experimental data based on assuming hadronlike behavior, $\sigma_{\gamma p} \sim
(\ln W^2)^2$\cite{PDG}. We recall that a dependence as 
$(\ln W^2)^2$ for hadron-hadron interactions was first predicted by Heisenberg
\cite{Heisenberg} and later recognized as the maximally
allowed growth by Froissart \cite{Froissart}  . We note that the hadronlike behavior of
photoproduction assumed in Fig.2 is associated with a $\ln W^2$ behavior  of the dipole cross section 
$\sigma^{(\infty)}(W^2)$, and not with the 
hadronlike $(\ln W^2)^2$ behavior corresponding to case iii).

The important conclusion from the above discussion is obvious. The measurement
of $\sigma_{\gamma^*p} (W^2,Q^2)$ at fixed $\eta (W^2,Q^2)$ as a function of
$W^2$, allows one to extract $\sigma^{(\infty)} (W^2)$, and, according to (\ref{eq:4.4}), allows
one to extract the $Q^2 = 0$ photoproduction cross section. Compare Fig.
\ref{fig:eta}, which illustrates the specific case ii) of $\sigma_{\gamma p}
(W^2) \sim (\ln W^2)^2$.

\begin{figure}
\includegraphics[width=8cm]{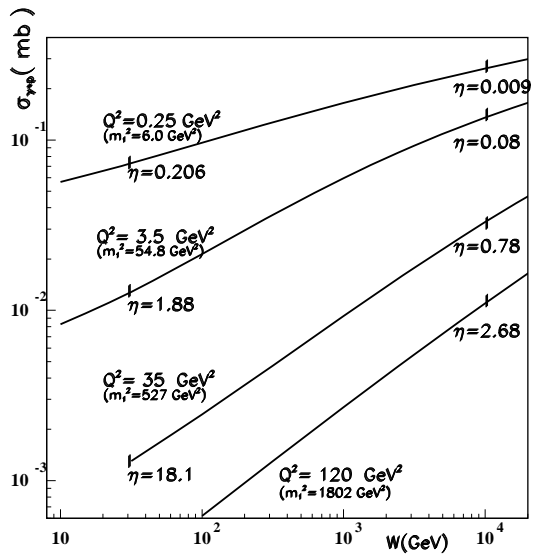}
\caption{The photoabsorption cross section as a function of the
energy $W$ for different values of $Q^2$. Note that a fixed
value of $Q^2$ is associated with a fixed mass range of
$q \bar q$ dipole states, $M^2_{q \bar q} \le m^2_1$ as determined
by (\ref{eq:4}). Compare also Table \ref{tab:tab2}. Transition
from color transparency to saturation at fixed $Q^2$ is a
consequence of the two-gluon coupling of the $q \bar q$ dipole
state.}
\label{fig:w_dep}
\end{figure}

In Fig. \ref{fig:w_dep}, we show the photoabsorption
cross section for fixed values of $Q^2>0$ as a function of $W$
reaching $W\cong 10^4$ GeV, the energy range discussed in 
connection with future electron-proton colliders\cite{Caldwell}. 
As indicated in Fig.3, fixed values of $Q^2$, according to 
(\ref{eq:4}), correspond to definite fixed values of 
$m_0^2\le M^2_{q\bar q}\le m_1^2$.
The approach to the true asymptotic limit \cite{NPB99, IJMPA31} of 
\be
\lim_{\substack{W \to \infty\\ Q^2_1, Q^2_2 > 0~{\rm fixed}}} 
\frac{\sigma_{\gamma^* p} (W^2,Q^2_1)}{\sigma_{\gamma^* p} (W^2, Q^2_2)}
= 1,
\label{eq:4.5}
\ee
or
\be
\lim_{\substack{W \to \infty\\Q^2 > 0~{\rm fixed}}} 
\frac{\sigma_{\gamma^* p} (W^2,Q^2)}{\sigma_{\gamma p} (W^2)}
= 1,
\label{eq:4.6}
\ee
according to Fig.3 is extremely slow.
Empirical evidence for the behavior in (\ref{eq:4.5}) and (\ref{eq:4.6})
can nevertheless be obtained from precise measurements at values of
$Q^2$ around $Q^2\cong 1$ GeV.

\section{Conclusions}
The present work is concerned with an interpretation of the 
photoabsorption cross section in terms of the range of the masses
$M_{q \bar q}$ of $\gamma^* \to q \bar q$ dipole states that
actively contribute to this cross section. The essential result
is contained in (\ref{eq:4}). The mass range of active
$q \bar q$ fluctuations is uniquely determined by a
proportionality to the photon virtuality $Q^2$. At fixed
$Q^2 \ge 0$, it is a fixed range of dipole masses that, as
a consequence of the two-gluon $q \bar q$ dipole coupling, with 
sufficient increase of the energy $W$ leads to the observed
transition from color transparency, $\sigma_{\gamma^*p}
(\eta (W^2,Q^2))\sim 1/\eta (W^2,Q^2)$ for $\eta (W^2,Q^2)\gg 1$, 
to saturation, $\sigma_{\gamma^*p} (\eta(W^2,Q^2)) \sim \ln
(1/\eta (W^2,Q^2))$ for $\eta (W^2,Q^2) \ll 1$. Alternatively,
at fixed energy $W$, a sufficient decrease in $Q^2$ towards
$Q^2 \cong 0$, associated with a decrease of the mass range
of active fluctuations, also leads from $\eta (W^2,Q^2) \gg 1$
(color transparency) to $\eta (W^2,Q^2) \ll 1$ (saturation).
Even though for $Q^2 > 0$ fixed, the active $q \bar q$ fluctuations
have a larger mass than at $Q^2 = 0$, in the true limit of
$W \to \infty$ the ratio of the cross section at fixed
$Q^2 > 0$, to the $Q^2 = 0$ photoproduction cross section
converges towards unity.

The low-x scaling of the photoabsorption cross section in
$\eta (W^2,Q^2)$ is weakly violated by a $\ln W^2$
dependence due to the dipole cross section, $\sigma^{(\infty)}
(W^2)$. The extraction of the W-dependence
of the dipole cross section from  DIS electron-proton scattering 
allows one to determine   the $Q^2 = 0$
photoproduction cross section and to verify or falsify
a hadronlike $(\ln W^2)^2$ dependence.

\begin{appendix}
\renewcommand{\theequation}{\Alph{section}.\arabic{equation}}
\section {The energy imbalance $\Delta E$.}
To make this paper self-confined, we add a  discussion on the
energy imbalance $\Delta E$ in (\ref{eq:1.1}).

Consider the transition of the (virtual spacelike) photon of virtuality $q^2=(q^0)^2-\vec q^{~2}=-Q^2\le 0$ 
to a $q\bar q$ state of four momentum  $K^\mu $ with
$K^2\equiv K^\mu K_\mu=(K^0)^2-\vec K^2>0$.  
With equality of the  three-momenta of the photon and the $q \bar q$ state, $\vec K=\vec q$,
the energy imbalance $\Delta E$ is given by
\bq
   \Delta E=K^0-q^0={{(K^0)^2-(q^0)^2} \over{K^0+q^0}}
             ={{Q^2+K^2}\over {K^0+q^0}}.
\label{eq:a.1}
\eq
We have to consider the high-energy limit of $q^0=|\vec q| \sqrt{1-{{Q^2}\over{\vec q^2}} }
\cong |\vec q|$ and $K^0=|\vec K|\sqrt{1+{{K^2}\over{\vec K^2}} }\cong |\vec K|=|\vec q|$,
where
\bqa
    \vec q^{~2} &\gg& Q^2,   \nonumber \\
    \vec q^{~2} &=&\vec K^2\gg K^2,
\label{eq:a.2}
\eqa    
and obtain
\bq
   \Delta E \cong {{Q^2+K^2}\over{2|\vec q|}}.
\label{eq:a.3}
\eq
To treat  the interaction of the photon with the proton of four-momentum $p_\mu$ and mass $M_p$, 
it is essential to  introduce the
center-of-mass energy squared, $W^2=(p+q)^2=M_p^2+2M_pq^0-Q^2$,
and  $q^0\equiv \nu$ in the proton rest frame,  and $x_{bj}\equiv {{Q^2}\over{2M_p\nu}}
\ll 1$, and accordingly also $W^2\simeq 2M_p\nu$.  The energy imbalance
(\ref{eq:a.2}) becomes
\bq
    \Delta E\simeq {{Q^2+K^2}\over{W^2}}M_p.
\label{eq:a.4}
\eq
It coincides with (\ref{eq:1.1}), since $M_{q\bar q}^2=K^2$, to the explicit representation
of which we turn now.

The four momenta of the quark and the antiquark are denoted by $k=(k^0,\vec k)$ and 
$k^\prime=(k^{\prime 0},\vec k^\prime)$, where $k^2=k^{\prime 2}=m_q^2$,
and, without  much loss of generality,  we assume massless quarks, $m_q=0$.
We choose  the $z$-axis of a coordinate system  in the direction of the three-momentum
$\vec q = \vec k+\vec k^\prime$.  For the ensuing discussion of the high-energy limit (\ref{eq:a.2}), 
it will be useful to represent the quark and antiquark momenta as
\bqa
    \vec k &=& z\vec q+\vec k_\perp, \nonumber \\
    \vec k^\prime &=& (1-z)\vec q -\vec k_\perp,
\label{eq:a.5}
\eqa
where  $\vec k_\perp\cdot \vec q=0$. 
The mass squared of the  $q\bar q$ state, $M_{q\bar q}^2$,  is given by 
\bqa
    M_{q\bar q}^2=K^2&=& (k^0+k^{\prime 0})^2-(\vec k+\vec k^{~\prime})^2  \nonumber \\
         &=& 2\sqrt{ (z(1-z)\vec q^2-\vec k_\perp^2 )^2 +\vec q^{~2}\vec k_\perp^2 } 
                \nonumber \\
       &&         -2(z(1-z)\vec q^2-\vec k_\perp^2)   \nonumber \\
        &=& 2\sqrt{ (k_zk_z^\prime-\vec k_\perp^2)^2+(k_z+k_z^\prime)^2\vec k_\perp^2}
                 \nonumber \\
       &&        -2(k_zk_z^\prime-\vec k_\perp^2),  
\label{eq:a.6}
\eqa        
where $k_z\equiv z|\vec q|$ and $k_z^\prime\equiv (1-z)|\vec q|$ were introduced 
in the last equality in (\ref{eq:a.6}).
One may check, as we did, that  (\ref{eq:a.6}) is (trivially) reproduced by applying 
a Lorentz transformation of magnitude $|\vec q|$   in the $z$ direction to the $q\bar q$ state at rest.

The relation (\ref{eq:1.1}) on $\Delta E$ requires finiteness of $K^2$ in the high-energy
limit of $z(1-z) \vec q^{~2}\gg \vec k_\perp^2$,  implying a necessary cancellation among
the $z(1-z)\vec q^{~2}$ terms in (\ref{eq:a.6}).   The cancellation occurs if and only if
$|z(1-z)|=z(1-z)$, or $z(1-z)>0$ or $0<z<1$.
Expansion of the square root in (\ref{eq:a.6}) for $z(1-z)\vec q^{~2}\gg \vec k_\perp^2$
yields
\bqa
  M_{q\bar q}^2=K^2&\simeq&2\bigl( z(1-z)\vec q^{~2}-\vec k_\perp^2\bigr)
              \nonumber \\
        &&  \times \Bigl( 1+{{\vec q^{~2}\vec k_\perp^2}\over{2(z(1-z)\vec q^{~2}-\vec k_\perp^2)^2}}\Bigr)
            \nonumber \\
          &&-2(z(1-z)\vec q^{~2}-\vec k_\perp^2),
\label{eq:a.7}
\eqa
or            
\bq
  M_{q\bar q}^2=K^2\simeq {{\vec k_\perp^2}\over{z(1-z)}}
        \Bigl( 1+{{\vec k_\perp^2}\over{z(1-z)\vec q^{~2}}}\Bigr)
           \simeq {{\vec k_\perp^2}\over{z(1-z)}}.
\label{eq:a.8}
\eq
           
We add the comment that upon solving the equation in (\ref{eq:a.6}) for $\vec k_\perp^2/z(1-z)$
in terms of $K^2$, $\vec q^{~2}$ and $z(1-z)$, one finds
\bq
    {{\vec k_\perp^{~2}}\over{z(1-z)}}={{K^2}\over{1+\frac{K^2}{\vec q^{~2}} }}
           \Bigl( 1+{{K^2}\over {4z(1-z)\vec q^{~2} }}\Bigr).
\label{eq:a.9}
\eq
Requiring $z(1-z)\vec q^{~2}\gg K^2$ reproduces the result (\ref{eq:a.8}).

From  (\ref{eq:a.9}), upon introducing $\sin^2\vartheta_{cm}$, where
$\vartheta_{cm}$ denotes the polar angle of the quark in the $q\bar q$ center-of-mass frame, 
\bq
      \sin^2\vartheta_{cm}={{\vec k_\perp^2}\over{\vec k_{cm}^2}} 
              ={{4\vec k_\perp^2}\over{K^2}},
\label{eq:a.10}
\eq
we find
\bq
     \sin^2\vartheta_{cm}={{4 k_\perp^2}\over{K^2}} = \frac {4z(1-z)+{{K^2}\over{|\vec q| ^2}} }
           {1+\frac{K^2}{\vec q^{~2}} }    \cong  4z(1-z).
\label{eq:a.11}
\eq
Combining (\ref{eq:a.10}) and (\ref{eq:a.8}) yields
\bq
    M_{q\bar q}^2 =K^2 = {{4\vec k_\perp^2}\over{ \sin^2\vartheta_{cm}}} \cong {{\vec k^2_\perp}\over{z(1-z)}}.
 \label{eq:a.12}
\eq  
The fraction $z$ of the momentum $\vec q$ of the photon taken over by the quark,
or rather the prduct $z(1-z)$,  in the $\vec q\to\infty$ linit yields the sine of the
polar angle $\vartheta_{vm}$.


In the CDP, we are exclusively dealing with the $\vec q^{~2} \to \infty$ limit, and
accordingly we replace the approximate  equalities in (\ref{eq:a.8}) and (\ref{eq:a.12})
by the equality    
\bq
    M_{q\bar q}^2=K^2  = {{4k_\perp^2}\over{\sin^2\vartheta_{cm}}} = {{\vec k_\perp^2}\over{z(1-z)}}.
\label{eq:a.13}
\eq
 This expression for the $q\bar q$ mass squared enters (\ref{eq:1.1}) and (\ref{eq:1.2})
 and all the subsequent considerations;  $M^2_{q\bar q}$ denotes the square of the
 $q\bar q$ mass in the $\gamma^* \to q\bar q$ transition to a $q\bar q$ state  with
 life time of order $1/\Delta E$.

\section{Comment on Saturation and Geometric Scaling.}

The representation of the experimental data in Fig. 1 for
$\xi = \xi_0 = 130$ in terms of the low-x scaling variable $\eta (W^2,Q^2)
= (Q^2 + m^2_0)/\Lambda^2_{sat} (W^2)$ looks similar to a
plot of the experimental data known as ``geometric scaling''\cite{Stasto}.
The result in \cite{Stasto} is a consequence of a ``saturation model'' \cite{Golec}
using an ansatz for the dipole cross section in the color-dipole
approach, $\hat{\sigma} (x,r) = \sigma_0 g (r/R_0(x))$, that depends
on Bjorken $x \cong Q^2/W^2$, and, accordingly, at any given energy
$W$ the dipole cross section depends on $Q^2$, in strong disagreement
with the very foundation of the color-dipole approach. The CDP
rests on the transition of the photon of spacelike virtuality,
$q^2 = - Q^2 < 0$, to massive $q \bar q$ states of timelike mass
squared, $M^2_{q \bar q} > 0$, associated with an energy imbalance
$\Delta E$ explicitly given in (\ref{eq:1.1}). The interaction of
the color-dipole-state  of mass $M_{q \bar q}$ with the gluon field in the 
proton depends on the $(q \bar q)p$ center-of-mass energy $W$
\cite{Sakurai}\cite{CSS}, in
no way different from e.g. $\pi p$ or $\rho^0 p$ interaction at
asymptotic energies,
and it cannot depend on the photon virtuality $Q^2$.
It must be concluded that the approach of the saturation model 
including its consequence of geometric scaling, even though leading
to a successful fit to the experimental results, suffers from employing $x \simeq
Q^2/W^2$ as argument of the dipole cross section, where $W^2$ should
be used, and it lacks a sound theoretical justification.

Color transparency and saturation,  in distinction from the "saturation model",
where "saturation" appears as an input assumption, in a consistent formulation of 
the CDP are recognized as a direct consequence of the two-gluon coupling
of the $q\bar q$-dipole states.  The relevance of the underlying energy 
imbalance $\Delta E$ between the spacelike photon of virtuality $q^2\equiv -Q^2<0$
and the timelike $q\bar q$ states of mass squared $M^2_{q\bar q}>0$, 
as pointed out in the main text, is quantitatively supported by
the experimental data.

\end{appendix}

\end{document}